 \documentclass[pmlr,twocolumn,10pt]{jmlr} 

\usepackage{booktabs}
\usepackage[load-configurations=version-1]{siunitx} 

\newcommand{\equal}[1]{{\hypersetup{linkcolor=black}\thanks{#1}}}

\theorembodyfont{\upshape}
\theoremheaderfont{\scshape}
\theorempostheader{:}
\theoremsep{\newline}

\jmlrvolume{LEAVE UNSET}
\jmlryear{2021}
\jmlrsubmitted{LEAVE UNSET}
\jmlrpublished{LEAVE UNSET}
\jmlrworkshop{Machine Learning for Health (ML4H) 2021} 

\usepackage{multirow}
\usepackage{multicol}

\title[Disparities in Dermatology AI: Assessments Using Diverse Clinical Images]{Disparities in Dermatology AI: Assessments Using Diverse Clinical Images}

\author{%
\Name{Roxana Daneshjou}\equal{These authors contributed equally} \Email{roxanad@stanford.edu}\\
\Name{Kailas Voldrahalli}\footnotemark[1] \Email{kailasv@stanford.edu}\\
\Name{Weixin Liang}\footnotemark[1] \Email{wxliang@stanford.edu}\\
\Name{Roberto Novoa} \Email{rnovoa@stanford.edu}\\
\Name{Melissa Jenkins} 
\Email{mjj@stanford.edu}\\
\addr Stanford University, CA, 94305\\
\Name{Veronica Rotemberg}
\Email{rotembev@mskcc.org}\\
\addr Memorial Sloan Kettering, NY, 10065\\
\Name{Justin Ko} 
\Email{jmko@stanford.edu}\\
\Name{Susan M Swetter}
\Email{sswetter@stanford.edu}\\
\Name{Elizabeth E Bailey}
\Email{ebailey2@stanford.edu}\\
\Name{Olivier Gevaert} \Email{olivier.gevaert@stanford.edu}\\
\Name{Pritam Mukherjee} \Email{pritamm@stanford.edu}\\
\Name{Michelle Phung} \Email{phungm@stanford.edu}\\
\Name{Kiana Yekrang} \Email{kyekrang@stanford.edu}\\
\Name{Bradley Fong} \Email{bafong88@stanford.edu}\\
\Name{Rachna Sahasrabudhe} \Email{rachnasahasrabudhe@gmail.com}\\
\Name{Albert Chiou}\equal{These authors contributed equally} \Email{achiou@stanford.edu}\\
\Name{James Zou}\footnotemark[2] \Email{jamesz@stanford.edu}\\
\addr Stanford University, CA, 94305
}

\begin{document}

\maketitle

\begin{abstract}
More than 3 billion people lack access to care for skin disease. AI diagnostic tools may aid in early skin cancer detection; however, most models have not been assessed on images of diverse skin tones or uncommon diseases. To address this, we curated the Diverse Dermatology Images (DDI) dataset—the first publicly available, pathologically confirmed images featuring diverse skin tones. We show that state-of-the-art dermatology AI models perform substantially worse on DDI, with ROC-AUC dropping 29-40 percent compared to the models’ original results. We find that dark skin tones and uncommon diseases, which are well represented in the DDI dataset, lead to performance drop-offs.  Additionally, we show that state-of-the-art robust training methods cannot correct for these biases without diverse training data. Our findings identify important weaknesses and biases in dermatology AI that need to be addressed to ensure reliable application to diverse patients and across all diseases.
\end{abstract}

\section{Introduction}
\label{sec:intro}

Globally an estimated 3 billion people have inadequate access to medical care for skin disease, and in the United States, there is a shortage of dermatologists \citep{coustasse-2019, tsang-2006}. Artificial intelligence (AI) tools have been suggested as a way to improve early skin cancer detection with rapid development of AI algorithms which claim to detect skin cancer over the last two years \citep{tschandl-2020, esteva-2017, daneshjou-2021}. There are currently no FDA approved AI devices for detecting skin cancer; however several commercial skin cancer detection applications have the CE mark in Europe \citep{freeman-2020}.

Systematic evaluation of state-of-the-art dermatology AI models on independent, real-world data has been limited. Most images used to train and test algorithms use private clinical image data with sparse descriptors \citep{daneshjou-2021}. Current publicly available data have limitations. The International Skin Imaging Collaboration (ISIC) dataset, which has pathologically confirmed dermoscopic images of cutaneous malignancies, lacks clinical images, images of inflammatory and uncommon diseases, or images of diverse skin tones \citep{tschandl-2018, codella-2019, kinyanjui-2020}. Online dermatology atlases, such as the Fitzpatrick 17k data set can have unreliable label sources with malignancies that are not pathologically confirmed \citep{groh-2021}.

We created the Diverse Dermatology Images (DDI) dataset to provide a publicly accessible dataset with diverse skin tones and pathologically confirmed diagnoses for AI development and testing. DDI enables model evaluation with stratification between light skin tones (Fitzpatrick skin types (FST) I-II) and dark skin tones (FST V-VI) and includes pathologically confirmed uncommon diseases (with incidence less than 1 in 10,000), which are usually lacking in AI datasets. We use this dataset to demonstrate three key issues for AI algorithms developed for detecting cutaneous malignancies:  1) significant drop-off in performance of AI algorithms developed from previously described data when benchmarked on DDI 2) skin tone and rarity of disease as contributors to performance drop-off in previously described algorithms and 3) the inability of state-of-the-art robust training methods to correct these biases without diverse training data.

\begin{figure*}[h]
\centering
\includegraphics[width=\textwidth]{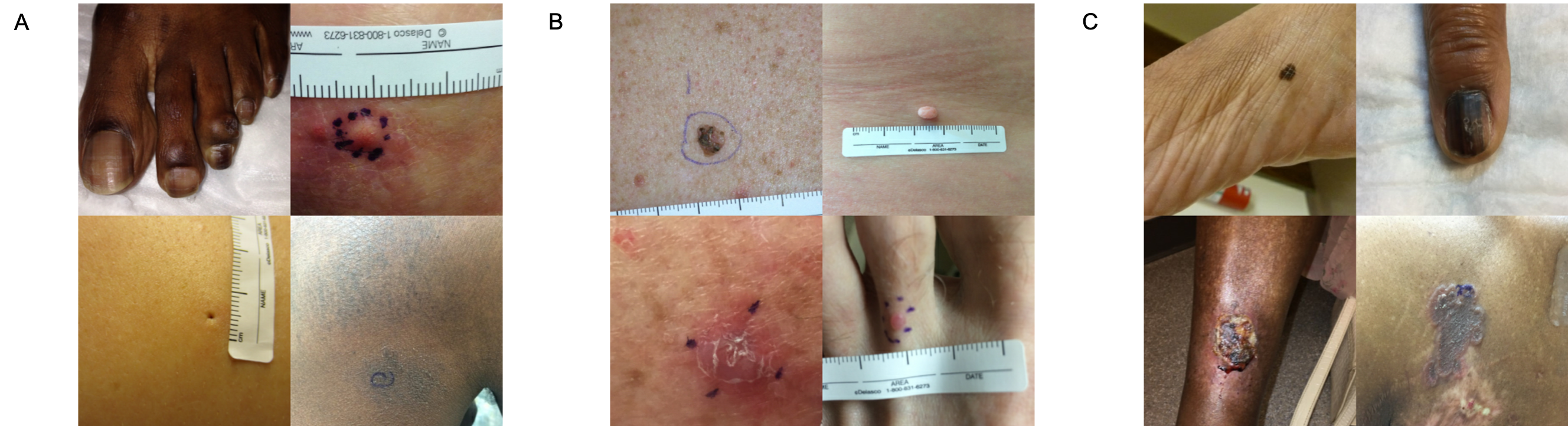}
\caption{
\small 
Examples of images from the entire DDI dataset for (A) all skin tones (B) FST-II and (C) FST V-VI.}
\label{fig:algs}
\end{figure*}

\section{Methods}
\label{sec:methods}

\subsection{Dataset design}
The DDI images were retrospectively selected by reviewing pathology reports in Stanford Clinics from 2010-2020 (Appendix A). Diagnosis labeling, image quality control, and FST labeling are described in Appendix A.

\subsection{AI Algorithms}

We tested 3 separate AI algorithms ( ``ModelDerm" and ``DeepDerm", and ``HAM 10000") trained to classify images as benign or malignant.

ModelDerm is a previously described algorithm \citep{han2020augmented}; an online API (https://jid2020.modelderm.com/) was used to generate outputs in December 2020.

DeepDerm is based on the algorithm developed by Esteva et al \citep{esteva-2017}. The algorithm uses the Inception V3 architecture \citep{szegedy2016rethinking}. Data sources include ISIC, two datasets sourced from Stanford hospital, Edinburgh Dermofit Library, and open-access data available on the web \citep{codella-2019, esteva-2017, ballerini2013color}. We mix all datasets and randomly generate train (80\%) and test splits (20\%). Images are resized and cropped to a size of 299 x 299 pixels. During training, we augment data by randomly rotating and selecting the largest inscribed upright rectangle and randomly vertically flipping images; we also use balanced sampling across benign and malignant images. DeepDerm is trained using binary cross entropy loss and Adam with a learning rate of $10^{-4}$ \citep{kingma2014adam}.

We also assess 3 robust training methods with DeepDerm's data: GroupDRO, CORAL, and CDANN \citep{sagawa2019distributionally, sun2016deep, li2018deep}. These methods require us to partition our dataset into the groups we would like to be robust across. Ideally, we would define these groups according to FST, but as we do not have this metadata, we use the source dataset as a proxy with the assumption that skin color distribution varies across data sources. As we have 5 source datasets and 2 classes (benign and malignant), we define 10 groups in total (5 sources x 2 classes).
Each of these robust models is trained using the same data augmentation strategies and optimization algorithm as described above, with the exception that we perform balanced sampling across these 10 groups and with a modified loss function according to the robust training method. 

HAM 10000 is an algorithm developed from the previously described, publicly available HAM 10000 dataset \citep{tschandl-2018}. The algorithm and training methods are identical to the DeepDerm methods, with the only difference being the data used. The HAM 10000 dataset consists of 10,015 dermoscopy images; all malignancies are biopsy confirmed.

\section{Results} 
\label{sec:evaluation}

\begin{table*}[htb]
\floatconts
  {tbl:algs}
  {\caption{
  \small 
  Performance (AUC-ROC and F1-Score) of algorithms on Diverse Dermatology Images dataset across skin color (all, FST I-II, and FST V-VI) and rarity of disease (all -- DDI and only common diseases -- DDI-C).}
    \vspace{-6mm}
  }%

  \centering
  {%
  \small
    \begin{tabular}{|l|l|l|l|l|l|l|l|}
    \hline
    \abovestrut{2.2ex}
    \multirow{2}{*}{\bf Algorithm} 
    & 
    \multirow{2}{*}{\bf Dataset} 
    & \multicolumn{3}{c|}{\bfseries ROC-AUC} & \multicolumn{3}{c|}{\bfseries F1-Score} \\ 
    \cline{3-8}
    \abovestrut{2.2ex}
    & & All & FST I-II & FST V-VI  
    & All & FST I-II & FST V-VI \\
    \hline
    \multirow{2}{*}{ModelDerm} & DDI   & 0.60 (0.54-0.66) & 0.64 (0.55-0.73) & 0.55 (0.46-0.64) 
    & 0.29 & 0.37 & 0.27  \\
   \cline{2-8} 
                               & DDI-C & 0.70  (0.63-0.77) & 0.68 (0.58-0.77) & 0.70 (0.57-0.82) 
    & 0.36 & 0.40 & 0.22 \\
    \hline
    \multirow{2}{*}{DeepDerm}  & DDI   & 0.53 (0.46-0.59) & 0.61 (0.50-0.71) & 0.50 (0.41-0.58) 
    & 0.30 & 0.37 & 0.20 \\
    \cline{2-8} 
                               & DDI-C & 0.62 (0.54-0.71) & 0.64 (0.54-0.75) & 0.55 (0.42-0.67) 
    & 0.29 & 0.37 & 0.15 \\
    \hline
    \multirow{2}{*}{HAM 10000} & DDI   & 0.65 (0.58-0.71) & 0.72 (0.63-0.79) & 0.57 (0.48-0.67) 
    & 0.08 & 0.04 & 0.11 \\
    \cline{2-8} 
                               & DDI-C & 0.71 (0.64-0.78) & 0.75 (0.66-0.82) & 0.625 (0.47-0.77)
    & 0.06 & 0.05 & 0.11 \\
    \hline
    \end{tabular}
  }
\end{table*}

\begin{table}[htb]
\small
\floatconts
  {tbl:fair_algs}
  {\caption{
  \small 
  Performance of robust training methods on FST I-II and FST V-VI. Mean (standard deviation) ROC-AUC values across 5 random seeds. ``DeepDerm" is copied from Table~\ref{tbl:algs}.}
   \vspace{-6mm}
  }%
  {%
  \small 
    \begin{tabular}{|l|l|l|l|}
    \hline
    \abovestrut{2.2ex}\bfseries Method & \bfseries Overall & \bfseries FST I-II  & \bfseries FST V-VI \\\hline
    \abovestrut{2.2ex}
    CDANN & 0.55 (0.03) & 0.60 (0.05)& 0.48 (0.03)\\
    CORAL & 0.54 (0.02) & 0.61 (0.04)& 0.45 (0.02)\\
    GroupDRO & 0.57 (0.02) & 0.62 (0.02)& 0.50 (0.02)\\\hline
    DeepDerm & 0.53 & 0.61 & 0.50\\\hline
    \end{tabular}
  }
\end{table}

After filtering poor quality images, there were 440 images representing 390 unique patients. The DDI dataset comprised a retrospective convenience sample to assess bias, so it was enriched for the lightest and darkest skin tones from the clinical database. There were a total of 208 images of FST I-II (159 benign, 49 malignant), 25 images of FST III-IV (24 benign, 1 malignant), and 207 images of FST V-VI (159 benign and 48 malignant) (Figure 1). There was no significant difference in photo quality scores between FST I-II photos and FST IV-V photos (Mann-Whitney U, p = 0.33).

We found algorithms had significant drop-offs compared to state-of-the-art performance on their own test sets. ModelDerm had a previously reported receiver operator curve area under the curve (ROC-AUC) of 0.93-0.94, while DeepDerm had a test set ROC-AUC of 0.88 and HAM 10000 had a test set ROC-AUC of 0.92. On the entire DDI dataset, ModelDerm had a ROC-AUC of 0.60 (95\% CI 0.54-0.66), DeepDerm had a ROC-AUC of 0.53 (0.46-0.59), and HAM 10000 had a ROC-AUC of 0.65 (95\% CI 0.58-0.71) (Table~\ref{tbl:algs}). Across all three algorithms on the DDI dataset, ROC-AUC performance was better in the subset of Fitzpatrick I-II images compared to Fitzpatrick V-VI with ModelDerm having a ROC-AUC of 0.64 (FST I-II) versus 0.55 (FST V-VI), DeepDerm having a ROC-AUC of 0.61 (FST I-II) versus 0.50 (FST V-VI), and HAM 10000 having a ROC-AUC of 0.72 (FST I-II) versus 0.57 (FST I-II). Performance gaps in ROC-AUC persisted between FST I-II and FST V-VI in the DDI common diseases for all algorithms except ModelDerm (Table 1).

To evaluate the robust training methods, each algorithm was run 5 times with different random seeds and results were averaged across seeds with mean and standard deviation in Table~\ref{tbl:fair_algs}. Though we note modest improvements in AUC value, the gap in performance between FST I-II and FST V-VI did not significantly improve with the robust training methods.

Because detecting malignancy is an important feature of these algorithms and clinical care, we assessed sensitivity to see if there was differential performance across skin tones. The thresholds for DeepDerm and HAM 10000 were determined by maximizing F1 score on the original test set. Thresholds for ModelDerm were previously reported \citep{han2020augmented}. Across the entire DDI dataset, two algorithms showed differential performance in detecting malignancies between FST I-II and FST V-VI: ModelDerm (sensitivity 0.41 vs 0.12, Fisher’s Exact Test p = 0.0025) and DeepDerm (sensitivity 0.69 vs 0.23, Fisher’s Exact Test p = 5.65 x $10^{-6}$).  HAM 10000 had poor sensitivity across all subsets of the data (sensitivity 0.04 in DDI and 0.03 in DDI-C) despite having a sensitivity of 0.68 on its own test set, suggesting that the threshold set for HAM10000 generalized poorly.

The data used to train ModelDerm, DeepDerm, and HAM 10000 come mostly from common malignancies, while the DDI includes uncommon diseases. To assess how this distribution shift could contribute to the model’s performance drop-off, we repeated this analysis in the DDI common diseases dataset. While removing uncommon diseases led to overall improvement in performance across all algorithms with ModelDerm having a ROC-AUC of 0.70 (CI 0.63-0.77), DeepDerm having a ROC-AUC of 0.62 (95\% CI 0.54-0.71) and HAM 10000 having a ROC-AUC of 0.71 (0.64-0.78), this performance was still lower than performance on the original test sets (Table~\ref{tbl:algs}). DeepDerm and HAM 10000 continued to have a disparity in performance between FST I-II and FST V-VI even among common diseases (Table 1). For DDI common diseases, thee disparity in sensitivity persisted for ModelDerm but was not statistically significant (sensitivity 0.41 to 0.28, Fisher’s Exact test, p=0.11) and remained significant for DeepDerm (sensitivity 0.71 vs 0.31, Fisher’s Exact test, p = 0.007).

\section{Discussion} 

To develop fair, generalizable AI in dermatology, datasets are required that include a diversity of skin tones and a range of diseases. The DDI dataset is a publicly available dataset of images including a diversity of skin tones, common and uncommon disease, with every lesion confirmed pathologically. Because benign lesions are not regularly biopsied, this dataset is enriched for “ambiguous” lesions that a clinician biopsied. While ambiguous lesions make the task of identifying benign versus malignant lesions more difficult, real world practice includes such lesions.

We find that previously developed algorithms have performance drop-offs on the DDI dataset compared to the ROC-AUC reported for their original test set. Uncommon diseases and dark skin tones contribute to this performance drop-off. DeepDerm training data included images from the same sites as the DDI dataset, thus, site-specific differences are less likely to play a role in differential performance. There are significant disparities in skin cancer diagnosis and outcomes; patients with dark skin tones get diagnosed at later stages, leading to increased morbidity and mortality ~\citep{sierro2021differences,agbai2014skin}. In order to alleviate disparities rather than exacerbate them, dermatology datasets used for AI training and testing must include diverse data.

An algorithm’s train and test data distribution influence its performance. ModelDerm was trained on a mix of Caucasian and Asian patients, while DeepDerm and HAM 10000 were trained on predominantly Caucasian patients. Unlike DeepDerm and HAM 10000, ModelDerm did not show a drop-off in ROC-AUC between Fitz I-II and Fitz V-VI on the DDI common skin diseases. The use of robust training methods to train a more unbiased version of DeepDerm did not improve disparity, suggesting that the limitations lie with the lack of diverse training data not the methods. HAM 10000 data has fewer diagnoses compared to DeepDerm or ModelDerm; however, every malignancy in HAM 10000 is pathologically confirmed, meaning less label noise~\citep{esteva-2017,tschandl-2018,han2020augmented}. The sensitivity-specificity threshold for all algorithms came from the original training/test data to avoid overfitting and to mimic how thresholds are set for commercial applications. These thresholds may not generalize to new datasets; for example, despite a good ROC-AUC, HAM 10000 had poor sensitivity on the DDI dataset, classifying most lesions as benign.

Limitations of this study include sample size; creating a dataset of diverse diagnoses across skin tones, where every lesion both benign and malignant has been biopsied requires significant expert curation. This iteration of the DDI dataset was meant to capture the effects of skin tone on performance, thus images from FST I-II and FST V-VI were enriched. Additionally, though FST is used most commonly for labeling skin tones for images used in AI studies, this scale has limitations and does not capture the full diversity of human skin tones~\citep{okoji2021equity}. Since most dermatology AI algorithms are not shared, we were limited in the number of algorithms we could test~\citep{daneshjou-2021}. Ongoing work includes expanding this dataset to provide more images for AI development and benchmarking in dermatology.


\bibliography{jmlr-sample}

\newpage
\appendix
\section{Data set design methodology}\label{apd:first}

An institutional database search tool, the Stanford Research Repository (STARR), was utilized to identify self-described Black patients who received a biopsy or tangential shave biopsy based on associated CPT code within the past 10 years starting from Jan 1, 2010.  Per institutional practice, all biopsied lesions undergo clinical photography taken from approximately 6 inches from the lesion of interest, typically using a clinic-issued smartphone camera.  These clinical images associated with the biopsied lesions were pulled from the Electronic Medical Record for analysis. Lesions were characterized by specific histopathologic diagnosis, benign vs. malignant status, date of biopsy, and age and gender of patient and Fitzpatrick skin type of the patient.

A control cohort composed of FST I/II patients was generated by matching each lesion diagnosed in the FST V/VI cohort with a lesion with the identical histopathologic diagnosis.  Patient demographics were assessed using chart review and included as a matched control if the lesion was diagnosed within the same 3 year period to account for incremental improvements in phone camera technology, and if the patient had the corresponding gender, age within 10 years, and FST type I/II.  In instances where a matching control strictly meeting the above criteria could not be identified, a lesion with a similar histopathologic diagnosis (i.e. matching one type of non-melanoma skin cancer with another type) in an FST I/II patient meeting the majority of the control criteria would be included instead to preserve the ratio of malignant to benign lesions.

Three independent board certified dermatologists were asked to assign a photo quality score using a previously developed image quality scale \citep{vodrahalli2020}.

Patient FST was confirmed by review of the clinical note, which was further verified against the demographic photo if available and the lesion image. All images underwent review by two board certified dermatologists (RD and AC) who used a consensus to resolve discrepancies. Light skin tones were FST I/II and dark skin tones were FST V/VI. FST III and IV were in the overall analysis but not enriched. 

Diagnosis labels and malignant versus benign labels were determined by a board certified dermatologist (RD) and board certified dermatopathologist (RN) reviewing the histopathologic diagnoses.  Any additional tests ordered on pathology was also considered for borderline cases – for example, an atypical compound melanocytic proliferation with features of an atypical spindle cell nevus of Reed was labeled as malignant due to diffuse loss of p16 on immunohistochemical stains, while an atypical lymphocytic infiltrate was confirmed to be benign based on negative clonality studies.

\section{Data and Code}\label{apd:second}
Data will be shared through the International Skin Image Collaboration (ISIC) repository.
Code for the algorithms can be found here:
\url{https://drive.google.com/drive/folders/1oQ53WH_Tp6rcLZjRp_-UBOQcMl-b1kkP}

\end{document}